\documentclass[10pt, a4paper]{iopart}

\usepackage[square, comma, sort&compress, numbers]{natbib}
\usepackage[usenames, dvipsnames]{color}
\usepackage{graphics}
\usepackage{epsfig}
\usepackage[caption=false, position=t, singlelinecheck=off]{subfig}
\usepackage{floatrow}
\usepackage{textcomp}
\usepackage{hyperref}
\usepackage{scalebar}
\usepackage[dvipsnames]{xcolor} 
\usepackage[tight]{units}
\usepackage{latexsym}
\usepackage{amssymb}

\hypersetup{colorlinks=true,
            linkcolor=red,
            urlcolor=blue,
            citecolor=Gray,
            plainpages=false}

\def\caion{$^{40}$Ca$^+$\,}

\begin{document}

\floatsetup[figure]{style=plain,subcapbesideposition=top}

\title{Operation of a planar-electrode ion-trap array with adjustable RF electrodes}

\author{M.~Kumph$^1$ P.~Holz$^1$ K.~Langer$^1$ M.~Meraner$^1$ M.~Niedermayr$^1$ M.~Brownnutt$^1$ R.~Blatt$^{1,2}$ }

\address{$^1$ Institut f\"{u}r Experimentalphysik,
Universit\"{a}t Innsbruck,
Technikerstrasse 25,
A-6020 Innsbruck, Austria}

\address{$^2$ Institut f\"{u}r Quantenoptik und Quanteninformation
der \"{O}sterreichischen Akademie der Wissenschaften,
Technikerstrasse 21a,
A-6020 Innsbruck, Austria}

\ead{muir.kumph@uibk.ac.at}

\begin{abstract}
One path to realizing systems of trapped atomic ions suitable for large-scale quantum computing and simulation is to create a two-dimensional array of ion traps. Interactions between nearest-neighbouring ions could then be turned on and off by tuning the ions' relative positions and frequencies. We demonstrate and characterize the operation of a planar-electrode ion-trap array. Driving the trap with a network of phase-locked radio-frequency (RF) resonators which provide independently variable voltage amplitudes we vary the position and motional frequency of a \caion ion in two dimensions within the trap array. With suitable miniaturization of the trap structure, this provides a viable architecture for large-scale quantum simulations.
\end{abstract}

\pacs{03.67.Ac  
    03.67.Lx,   
    05.60.Gg,   
    37.10.Ty}   
\vspace{2pc}

\maketitle
\renewcommand{\thefootnote}{\arabic{footnote}}

\section{Introduction}

Trapped atomic ions continue to make advances towards the realization of powerful quantum information processors \cite{Blatt:2012, Monroe:2013}. There exist several distinct proposals for scaling the current proof-of-principle demonstrations to systems with hundreds or thousands of ionic qubits. These include using a single, anharmonic trap with a long chain of ions \cite{Lin:2009}; creating an array of traps through which small groups of ions can be shuttled \cite{Kielpinski:2002}; and generating an array of separate traps, between which interactions can take place \cite{Cirac:2000, Bermudez:2012}. The latter method of scalability lends itself well to both one-dimensional (1D) and two-dimensional (2D) arrays of ions. 2D arrays could be particularly useful for realizing simulations of two-dimensional systems \cite{Hauke:2013, Shi:2013}; for creating entanglement resources for measurement-based quantum computing \cite{Briegel:2009}; and for facilitating robust quantum computation architectures \cite{Nigg:2014}.

In pursuing an architecture consisting of a 2D array of trapped ions we consider a scheme in which ions within the array interact with their nearest neighbours. The tunable interaction strengths can be turned on and off using such schemes as proposed by Cirac and Zoller \cite{Cirac:2000} and Bermudez et al. \cite{Bermudez:2012}, and successfully implemented in one dimension to entangle ions in separate wells \cite{Wilson:2014}. In these schemes the time required to implement a gate between a pair of ions (each of mass $m$ and charge $q$) on resonance scales with the inter-ion separation, $a$, and the ions' motional frequency, $\omega$, as
\begin{equation}
\label{eq:GateTime}
T_{\rm gate}=\frac{4 {\rm \pi} \epsilon_0 m a^3 \omega}{q^2}.
\end{equation}
Consequently, tuning the interactions requires control over the separation of ions in the two wells, and over the ions' motional frequencies.

Along one axis of a linear trap the ions are confined by static potentials. Consequently the relative positions and frequencies of trapped ions along this direction can be tuned by varying DC voltages \cite{Harlander:2011, Brown:2011_03}. By this means, ions in separate wells have been entangled \cite{Wilson:2014}. The situation in a 2D array of traps, however, is more complicated as the ions are confined in at least two (and possibly three) directions by a pseudopotential created by a radio-frequency (RF) field. Consequently, variation of the inter-ion spacing---while keeping the ions at their respective RF nulls---requires control of variable RF voltages.

There have been experimental demonstrations of confining ions in a 2D array of ion traps \cite{Clark:2009} and hopping ions between sites in a miniaturized array of traps \cite{Sterling:2014}. However, as the RF potential could not be locally varied, it was not possible in these experiments to move the positions of the RF nulls. Chiaverini and Lybarger \cite{Chiaverini:2008, Lybarger:2010PhD} have proposed dynamically re-configurable arrays in which RF electrode ``pixels" could be switched on or off. We have proposed a modification to the basic 2D array architecture by which the RF electrodes are segmented, allowing the positions and frequencies of trapped ions in a 2D array to be tuned by varying RF voltages \cite{Kumph:2011}. As the RF voltage on a particular electrode is reduced, the electric field above that electrode falls, and the ion moves towards the region of reduced electric field. This is then analogous to the tuning of ions' positions and frequencies in a 1D array by varying DC voltages. 

There are a number of methods which may be used to apply RF voltages of different amplitudes to different electrodes, and thereby control the position of the RF null. At low frequencies, dust traps holding lycopodium spores have been reconfigured using variable alternating voltages of 50~Hz \cite{Kumph:2011}. In ion traps, different electrodes can be driven with different amplitude voltages by making different tapping points in a single helical resonator \cite{VanDevender:2010}. The RF null has also been moved in a more smoothly continuous fashion by selectively adjusting the load capacitance of the trap electrodes \cite{Herskind:2009, Kim:2010}. This latter method has been used to switch between trapping configurations in which ions were in a single linear pseudopotential minimum and in two separated linear traps \cite{Tanaka:2014}.

Variation of the capacitive load can result in different RF electrodes having different phases, leading to significant micromotion. The method used in the present work is able to vary the amplitude of the RF while actively keeping the RF phase constant by feeding back to a varactor diode. This allows the RF minimum to be significantly displaced within microseconds (the speed being limited by the trap frequency), while ensuring that there is minimal phase-induced excess micromotion. We implement a system of multiple low-power RF drives with actively locked frequencies, for which the voltage amplitudes can be independently varied in real time. These are used to trap single \caion ions and vary the ion's position in two dimensions within a 2D array of traps, as well as varying its motional frequency.

The trap apparatus, including the associated driving electronics, is described in section~\ref{sec:Setup}. Trapping results, including the variation of an ion's position and motional frequency as a function of the addressable RF voltage, are reported in section~\ref{sec:Results}. To realize a large-scale quantum information processor, the building blocks demonstrated here must be implemented in a system with $\gtrsim$100 ions and $\lesssim$100\,\textmu m inter-well separation; section~\ref{sec:DiscussionAndOutlook} discusses the path to realizing such a system.

\section{Setup}
\label{sec:Setup}

The apparatus used for the demonstration of controllably varying a trapped atomic ion's position and motional frequency within a 2D array of traps has two distinct aspects to consider: the trap array and the drive electronics. A new trap array was fabricated using PCB technology. This fabrication method is simple and reliable, but limits the trap dimensions to be larger than those required for coherent quantum operations. Importantly, however, it permits for the first time the demonstration of the variable RF drive electronics, operating in a parameter regime congruent with scalable trapped-ion quantum computing. The setup of the trap (including the attendant mounting and vacuum system) and the driving electronics are discussed here in turn.

\subsection{Ion-trap array}
\label{sec:trapfab}

\begin{figure*}[t]
\setlength\unitlength{10pt}
\begin{picture}(2,2) 
\thicklines
\put(0,0){\circle*{.3}}
\put(0,0){\vector(1,0){2}}
\put(0,0){\vector(0,1){2}}
\thinlines
\put(-.4, -.4){\makebox(0,0)[cb]{\small z}}
\put(2,0.25){\makebox(0,0)[cb]{\small x}}
\put(0,2){\makebox(0,0)[cr]{\small y}}
\end{picture}\quad
\includegraphics[width=10cm]{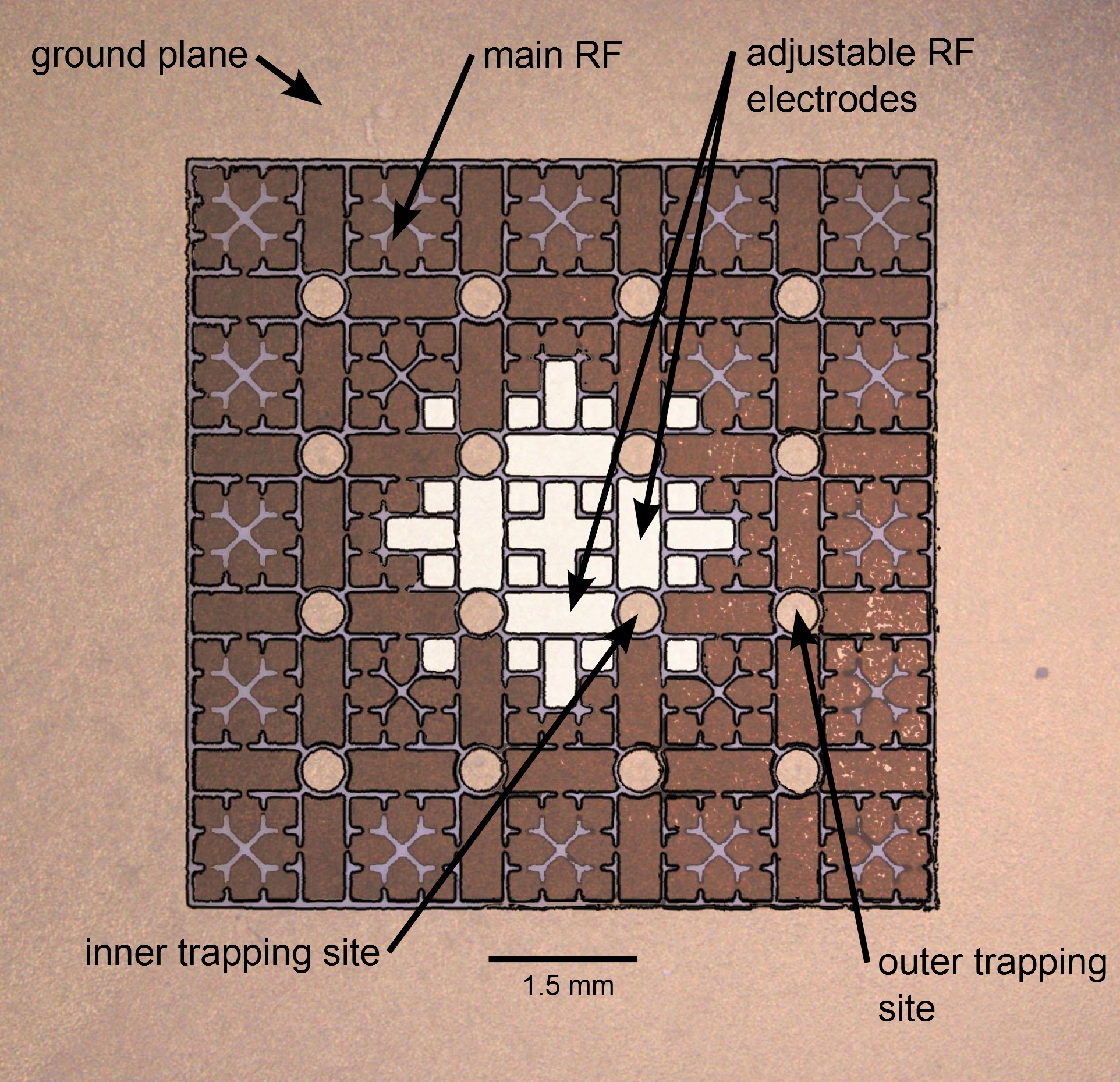}\\
\caption{False-colour image of Folsom.  Circular trapping-site electrodes are held at RF ground, and ions can be trapped 400~\textmu m above each of these.  The outer 12 trapping-site electrodes are connected to DC ground, while the inner 4 sites have separate connections to allow non-zero DC voltages to be applied.  The large square outer electrode is held at ground.  There are two types of RF electrodes:  The main RF electrode (darkened in figure) forms a single electrode and allows trapping at the outer 12 trapping sites. The inner RF electrodes (lightened in figure) are connected individually to allow independent adjustment of the RF voltages applied. All electrodes are connected by vias to traces on the backside of the printed circuit board. Experiments were carried out first at an outer trapping site and then in a different array at an inner trapping site.}
\label{fig:wholearray}
\end{figure*}
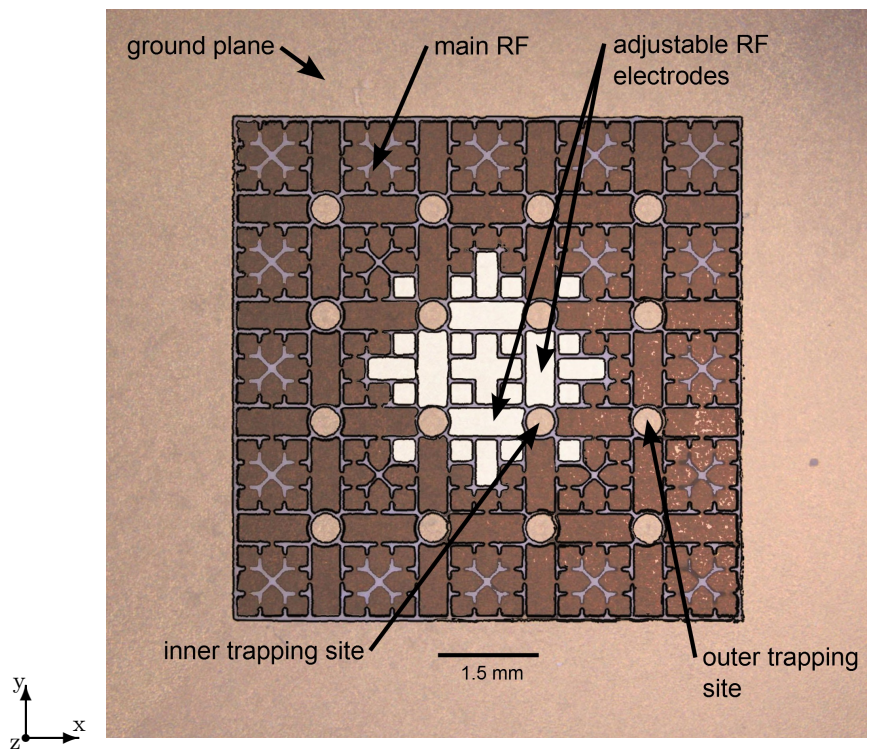

\begin{figure*}[t]
\subfloat[\hspace{2mm}\small\scalebar{1.8cm}{1}{2}{0}{10}{mm}\normalsize\label{fig:folsompcb}]{
\includegraphics[width=6.25cm]{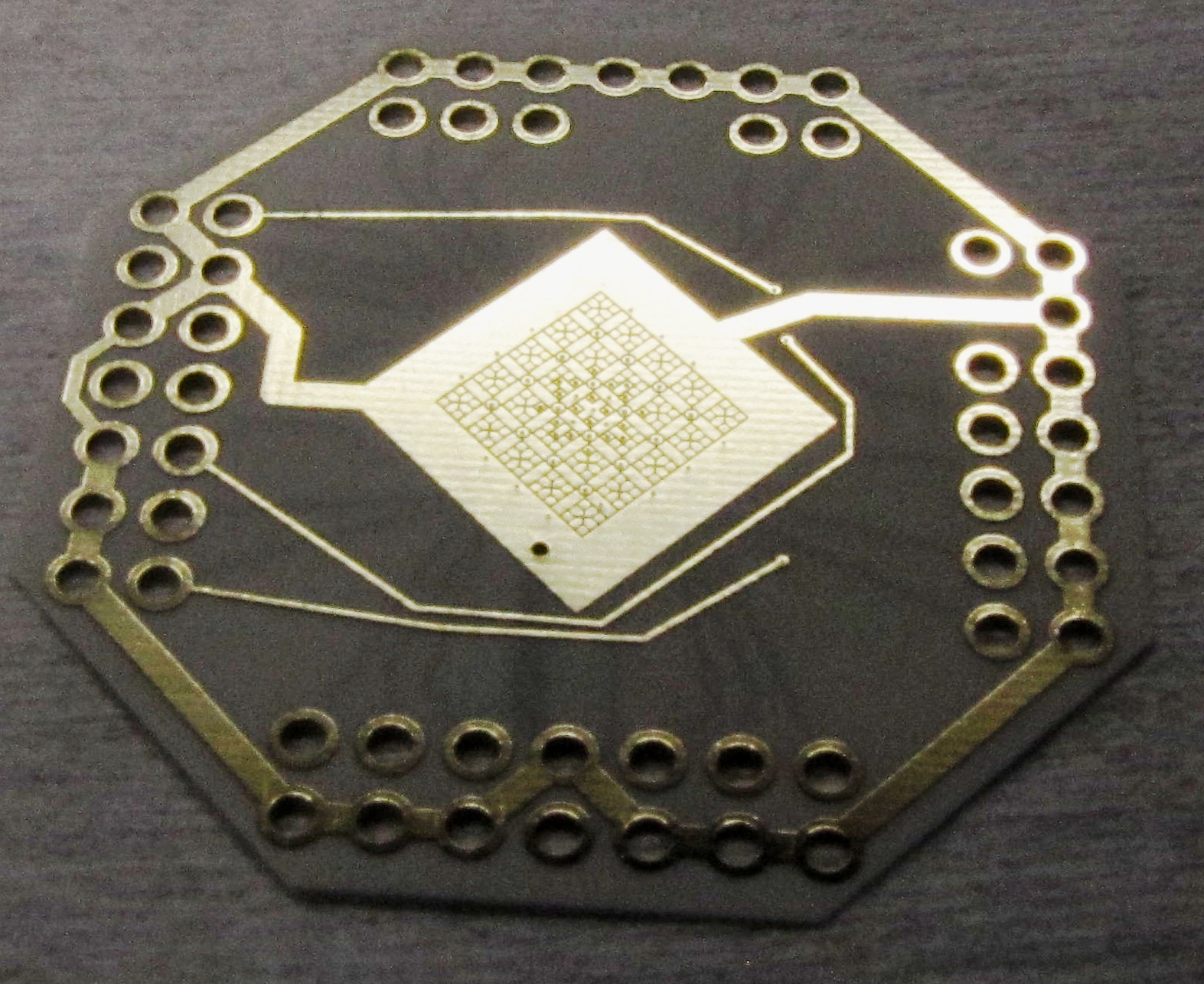}
}
\subfloat[\hspace{2mm}\small\scalebar{1.35cm}{1}{2}{0}{500}{\textmu m}\normalsize\label{fig:folsomdetail}]{
\includegraphics[width=6.25cm]{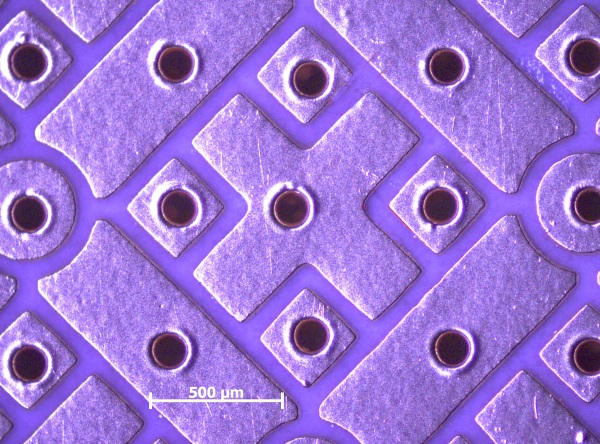}
}\\
\subfloat[\hspace{2mm}\small\scalebar{1.8cm}{1}{2}{0}{10}{mm}\normalsize\label{fig:folsomconnectorized}]{
\includegraphics [width=6.25cm]{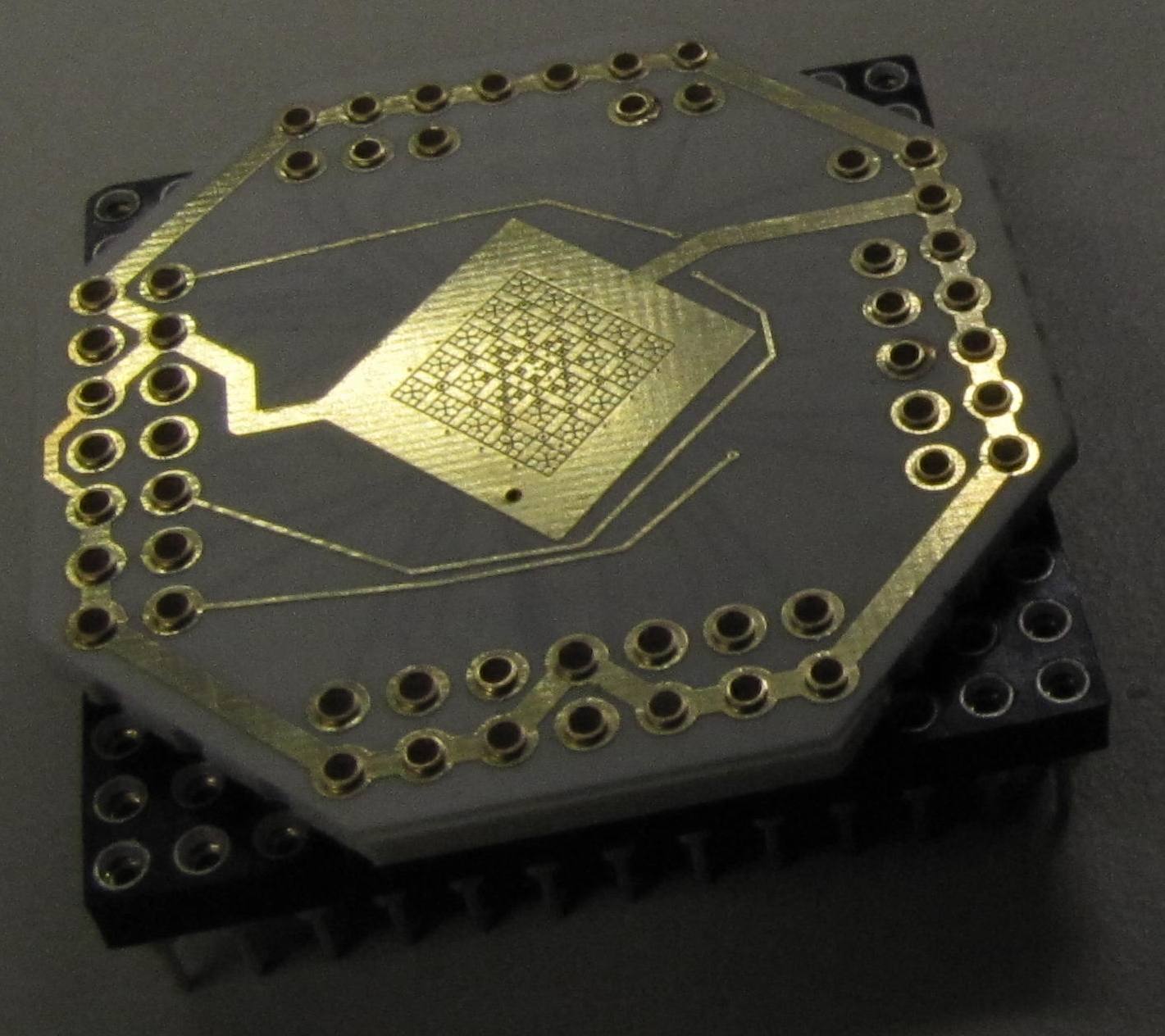}
}
\subfloat[\hspace{2mm}\small\scalebar{1.7cm}{1}{2}{0}{2}{cm}\normalsize\label{fig:folsomwiredup}]{
\includegraphics [width=6.25cm]{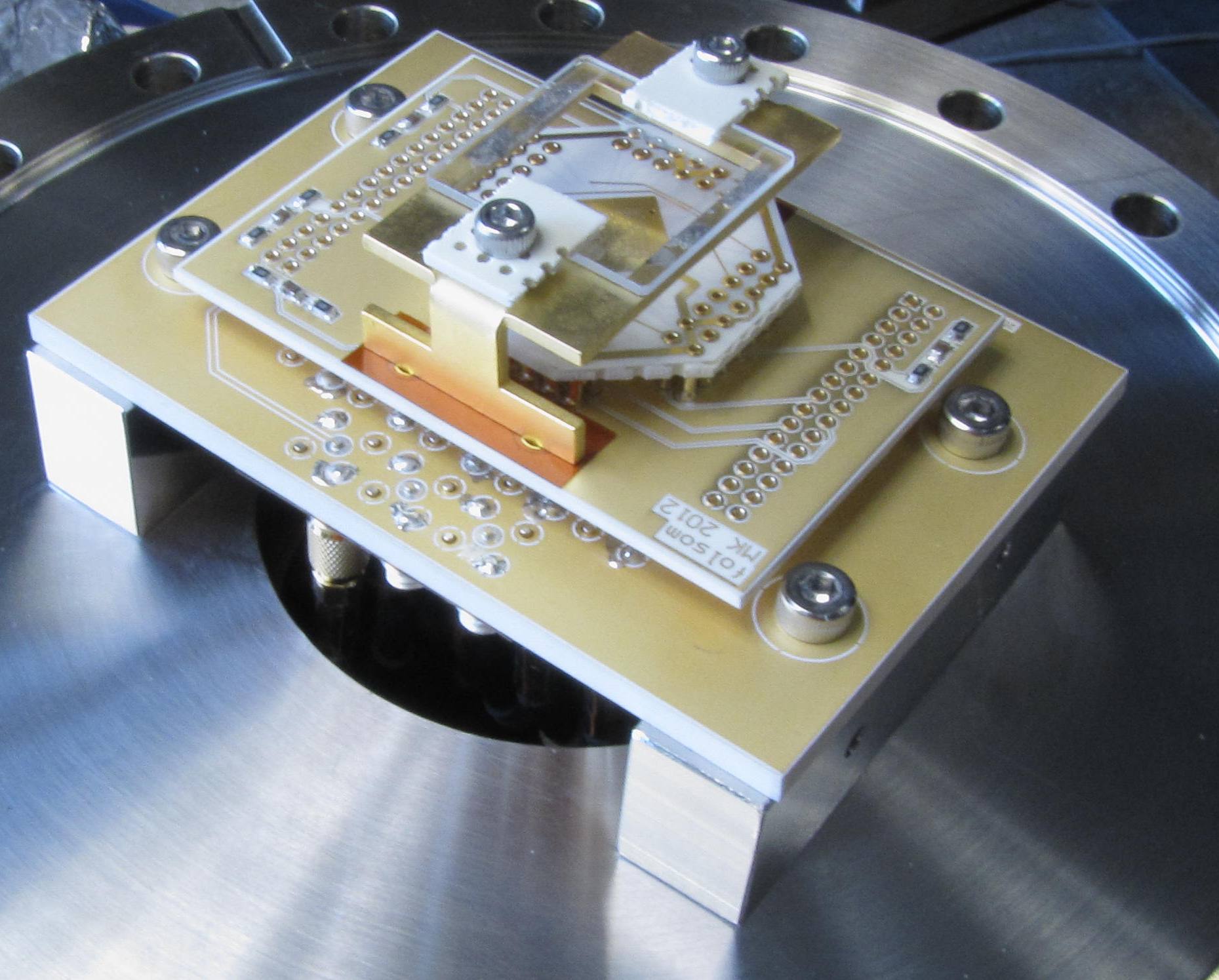}
}
\caption{Fabrication and mounting of the ion trap array.  (a) Printed circuit board after plating with gold. (b) Detail of the centre electrodes' structure. The black circles are holes where vias connect the electrodes to the backside of the PCB. (c) Connectorized trap. Electrical connections from the trap to the socket are made with pins crimped to the vias at the edge of the PCB. (d) Trap mounted on an in-vacuum filter board and cable breakout board. A clear conductive window is mounted 7\,mm from the trap surface. When the entire assembly is installed in the vacuum chamber the trap faces downwards.}
\label{fig:folsomfab}
\end{figure*}

Following a tradition of naming traps after famous prisons, the trap array design considered here is dubbed ``\textit{Folsom}". The trap array's electrode structure is shown in figure \ref{fig:wholearray}.  There are 16 circular trapping-site electrodes, each of which is held at RF ground. Above each of these a pseudopotential minimum is formed, 400~\textmu m above the trap surface, which serves to confine ions in all three dimensions,  The principal axes of the ions' motion are aligned along the $x$, $y$, and $z$ directions shown in figure~\ref{fig:wholearray}. From symmetry considerations, motion in the $xy$ plane is termed radial motion, and that in the $z$ direction is termed axial motion. Between the trapping-site electrodes of the inner $2 \times2$ array there are individually adjustable RF electrodes. These allow the separation and motional frequency of nearest-neighbouring ions to be reduced, as is required for tunable interactions [see equation\,(\ref{eq:GateTime})]. The size of the electrode features is set by the fabrication method used. The resulting trap separation of 1.5~mm is too large to allow coherent quantum operations between the trapping sites \cite{Kumph:2011}. Nonetheless, the structures are sufficiently small to permit a demonstration of the electronic control needed for future, smaller trap arrays. The outer 12 trapping-site electrodes are surrounded by a single planar RF electrode. While their potentials cannot be individually adjusted, the outer sites provide realistic boundary conditions for the inner sites, which will ultimately need to form part of a larger array. A ground plane provides the final electrode needed to generate an RF quadrupole trap. An additional far-field ground is provided by a transparent ground plane (shown in figure~\ref{fig:folsomfab}d) mounted 7~mm from the plane of the trap.  This has the effect of increasing the trap depth and shielding the trap from stray fields. In the present design this plane is grounded though, if necessary, a DC connection could be made so that it provides a uniform static field, such as may be of use for micromotion compensation.

The steps for Folsom's assembly are shown in figure~\ref{fig:folsomfab}. The electrode structure was etched\footnote{fabrication by Andus Electronic GmbH} into the 18~\textmu m thick copper cladding of a printed circuit board (PCB)\footnote{Rogers 4350b substrate ($35\,{\rm mm} \times 35{\rm mm} \times 170$~\textmu m)}. The copper was subsequently plated with 10~\textmu m of gold using a gold sulphite solution\footnote{Transene TSG-250}.  The trap after gold plating is shown in figure~\ref{fig:folsompcb}, with the surface detail shown in figure~\ref{fig:folsomdetail}. The board was then mounted to a 1.6~mm thick substrate using pins\footnote{Mill-Max 3116} which were crimped into vias in the PCB, shown in figure~\ref{fig:folsomconnectorized}. This served both to support the PCB and to connectorize it. The connectorized trap and a ground plane\footnote{Indium tin oxide (70-100 $\Omega/\Box$) on a fused-silica substrate (25~mm $\times$ 25~mm $\times$ 1~mm)} were mounted on a custom-made PCB filter board, which was in turn plugged into a cable breakout board. The completed trap assembly is shown in figure~\ref{fig:folsomwiredup}. Coaxial cables from the breakout board were connected to the coaxial feedthrough described below.

\begin{figure*}[t]
\includegraphics[width=8cm]{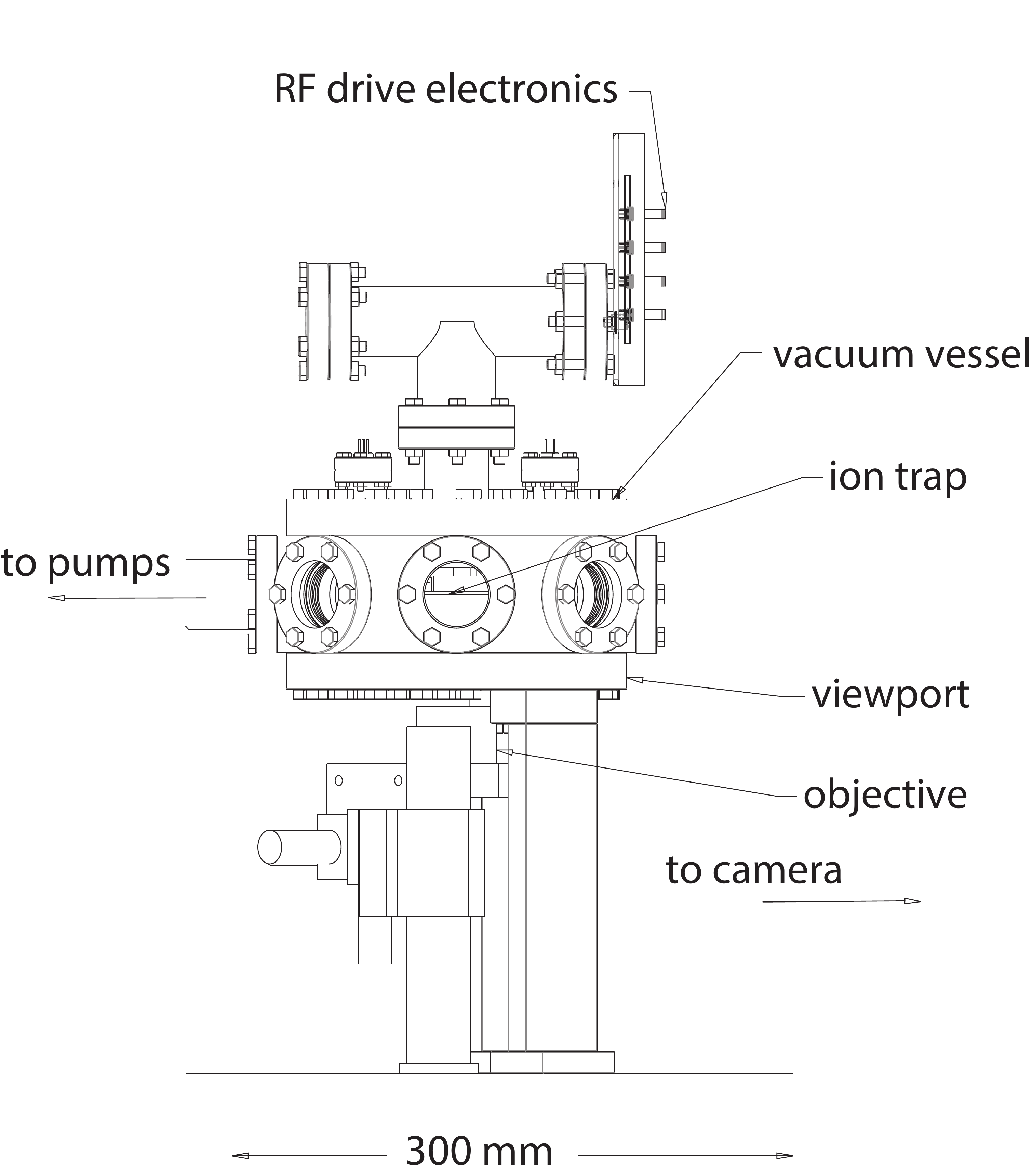}
\caption{Vacuum setup for testing Folsom. RF feedthroughs at the top allow the trap to be driven by the RF electronics. Imaging of the ions is done through a viewport on the bottom of the vacuum vessel, through an objective. Lasers for cooling and photoionization have access through the viewports on the sides of the vacuum vessel.}
\label{fig:folsomsetup}
\end{figure*}

The vacuum configuration for the experiment is shown in figure~\ref{fig:folsomsetup}. The trap chip was held in an octagonal vacuum chamber at a pressure of $10^{-10}$~mbar with the trap surface oriented horizontally and facing downwards. The chamber had coaxial feedthroughs mounted to the top flange which provided electrical connections from the RF electronics to the individual trap electrodes. A viewport on the bottom flange allowed imaging of the ions onto a CCD camera (not shown in the figure) by a custom objective. The viewports on the sides of the vacuum chamber provided laser access in the plane of the trap array. The ions were cooled and imaged with laser light of wavelength 397~nm and 866~nm \cite{Urabe:1992}.

\subsection{Variable RF sources}
\label{sec:rfsource}

\begin{figure*}[htb]
\includegraphics[width=12cm]{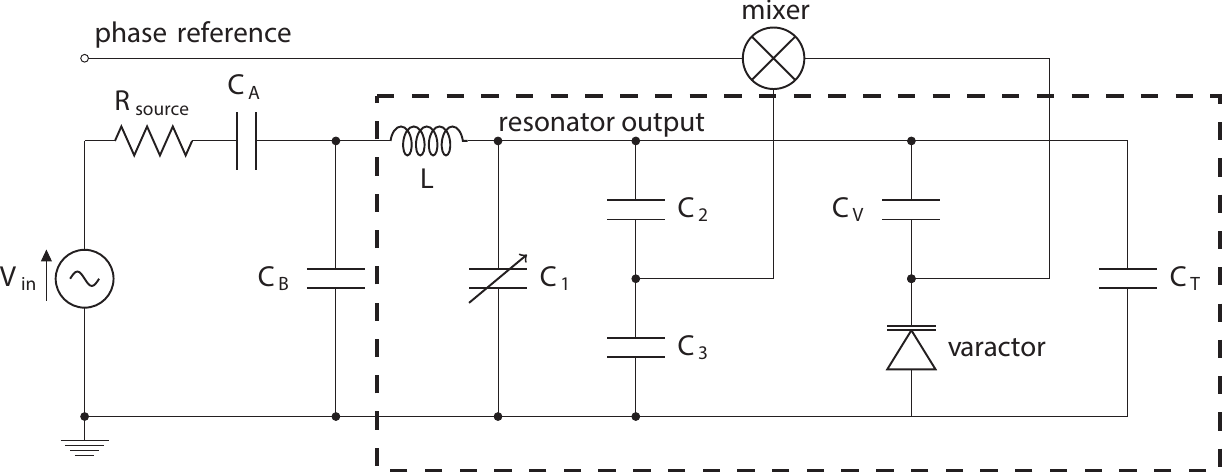}
\caption{Schematic of the phase-locked RF resonator.  A tank resonator is formed by the high-quality inductor, L, and the capacitors shown inside the dashed box. The mixer locks the output phase of the resonator to a phase reference by varying the capacitance of the varactor diode.}
\label{fig:phaselock}
\end{figure*}

Controlling the position and motional frequencies of a specific ion within an array requires the relative amplitudes of the RF voltages on adjacent electrodes to be varied. Any relative phase between the voltages on different electrodes, however, will cause the ion to undergo excess micromotion \cite{Berkeland:1998, Herskind:2009}, and so it is required that all RF voltages remain in phase. To this end a novel RF drive was developed, a simplified schematic of which is shown in figure~\ref{fig:phaselock}. In essence, the drive uses a tank resonator to match the impedance of the trap electrode and wiring to that of the RF source, while also using a control circuit to lock the phase of the RF drive's output to a reference. Each RF electrode is then individually driven by such an independently phase-locked RF drive.

When the voltage amplitudes on different electrodes are varied, the capacitance, $C_{\rm T}$, of a particular trap electrode to ground may also change. The control circuit shown in figure\,\ref{fig:phaselock} compares the output voltage of the resonator to a phase reference and adjusts the capacitance of the varactor diode to compensate for any changes in $C_\mathrm{T}$. The tank resonator itself is formed by an inductor, L, and the capacitors shown inside the dashed box in figure~\ref{fig:phaselock}. The variable capacitor, $\mathrm{C_1}$, allows the frequency of the resonator to be tuned. Capacitors $\mathrm{C_2}$ and $\mathrm{C_3}$ form a voltage divider to provide a low-voltage pick-off which has the phase of the RF drive's output. Capacitor $\mathrm{C_V}$ protects the varactor diode from the RF drive's high output voltage. The capacitors $\mathrm{C_A}$ and $\mathrm{C_B}$ match the impedance of the tank resonator to the 50~$\Omega$ RF source.

In the experiments described below, the voltages on two (of a possible 25) RF electrodes are independently varied. These electrodes have only a light capacitive load (30~pF) and require the phase to be actively locked in order to keep the phase-walk below $\sim$1$^\circ$. The remaining RF electrodes are bundled into groups and driven by three separate RF resonators, which are adjusted to have the same phase. The total capacitive load on each of these resonators is around 200~pF, meaning that the relative changes in the load due to mutual capacitance are sufficiently small to obviate the need to lock their phases.

Correct adjustment of the system requires that the source (V$_{\rm in}$) has the correct phase and amplitude, that the resonator has the correct resonant frequency (set by C$_{1}$), and that the source and resonator are impedance matched (set by C$_{\rm A}$ and C$_{\rm B}$). To achieve this the RF electrodes are all initially set to have nominally the same voltage amplitude, frequency and phase. In this ``home" configuration the optimal conditions are met when the RF power reflected from each of the tank resonators (measured using directional couplers) is minimized.

Changing an ion's position and motional frequency requires the amplitude of one of the RF voltages to be reduced, without altering its frequency or phase. Under such circumstances some amount of RF power is coupled capacitively between RF electrodes with different voltage amplitudes, causing power to be transmitted back up the line, through the resonator, to the RF source. For this reason the method of minimizing the signal through a directional coupler cannot be used away from the home settings to check that the circuit is operating as it should. Instead, an oscilloscope with capacitive-pickup probes is used as an out-of-loop diagnostic to monitor the output phase and amplitude of each RF drive.

\section{Results}
\label{sec:Results}
The operation of the Folsom array has been characterized. This section first presents the trapping of single ions at one of the outer 12 trapping sites of the array using a single RF source and ground. Secondly, trap operation at one of the inner 4 trapping sites using multiple phase-locked RF sources is presented. At the inner site the voltages on specific RF electrodes can be individually and independently adjusted to demonstrate control over both the ion's position and its motional frequency in two dimensions. Finally, heating-rate measurements made at the outer and the inner trapping sites are presented.

\subsection{Ion Trapping}
\label{subsec:InitialTrapping}
Single \caion ions were loaded at one of the outer trapping sites (indicated in figure~\ref{fig:wholearray}) using a single RF trap drive connected to the main RF electrode. The ion's motional frequency in the $z$ direction (normal to the trap surface) was measured to be 680~kHz at a trap-drive frequency of 10.7~MHz and an RF voltage amplitude of 100~V (0-pk). The lifetime of a single ion with the cooling lasers on was approximately 10 minutes.

The trapping behaviour at the outer trapping site deteriorated over time, with increased difficulty in loading, increased micromotion, and decreased trapping lifetimes. Possible reasons for this behaviour are discussed below. After several months of such deterioration it was no longer possible to load ions. A new trap array (of the same design) was then installed and used for all subsequent experiments described in this paper.

In the new trap array, single \caion ions were loaded at one of the inner trapping sites (indicated in figure~\ref{fig:wholearray}). For this, multiple RF drives were used to apply RF voltages of the same phase but independently variable amplitudes (see section~\ref{sec:Setup}) to the various adjustable RF electrodes. Several of the segmented RF electrodes were connected together, so that five separate RF drives were used to drive the electrodes surrounding the trapping-site. Despite the increased complexity of the drive electronics the inner trapping site in this array exhibited similar uncooled lifetimes to the outer trapping site in the first trap array tested.

The home configuration of the multiple RF drives was set such that all of the RF electrodes had the same RF voltage amplitude and phase, mimicking one single, unsegmented RF electrode.  The RF drive voltage had an amplitude of 100~V and a frequency of 10.1~MHz. The voltage at the individual electrodes was then varied by changing the power applied to the relevant RF resonator.  All RF power data below is given in dB relative to the above-specified home configuration.

\begin{figure}[t]
\includegraphics[width=8cm]{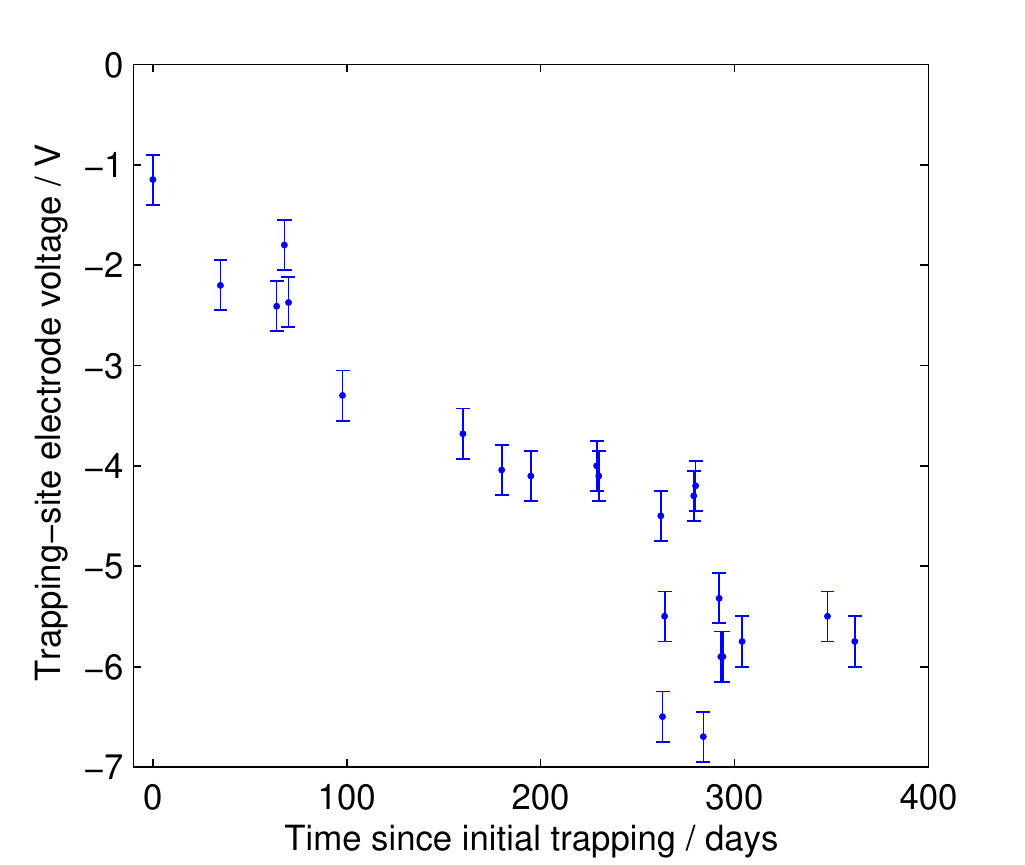}
\caption{The DC voltage that had to be applied to the trapping-site electrode to allow stable trapping. This became more strongly negative over the course of a year, until it was no longer possible to trap ions.}
\label{fig:trappingvoltage}
\end{figure}

Trapping ions in the second trap array required that the trapping-site electrode beneath the ion be held at a negative voltage. The voltage required was initially around -1~V but, over the course of one year, became more strongly negative so that ultimately a voltage of around -6~V was necessary. The progression of the required DC voltage over time is shown in figure~\ref{fig:trappingvoltage}. At the outer trapping sites (as used in first trap array) the trapping-site electrode was wired directly to ground, and so it was not possible to apply non-zero voltages. It may be that the observed deterioration in the ability to trap ions at an outer trapping site (described above) is related to the need for an offset voltage at the inner trapping site described here.

Application of large DC voltages to the circular trap site electrode complicates the minimization of micromotion when the ion is away from the home position. It will therefore be important to minimize or eliminate the cause of this large, increasing, negative voltage. While this experiment is not unique in observing static compensation voltages which drift significantly over time \cite{Brownnutt:2007, Haerter:2014, Narayanan:2011} the source of the problem in the situation at hand is not currently known. It may have been caused by a build up of calcium deposited from the oven onto the trap electrodes. This problem can be mitigated by better shielding of the trap electrodes from the oven \cite{Britton:2009} or by use of a separate loading zone \cite{Britton:2009, Blakestad:2009}. Alternatively the problem may have been due to materials issues in the electrodes, such as the copper of the electrode bulk diffusing through the gold plating \cite{Pinnel:1972, Tompkins:1976} or the growth of field-emission tips \cite{Steinhauer:2011}. These issues could be overcome by a different selection of materials from which to fabricate the trap.

\subsection{RF control of ion position}
The proposed implementation of tuning the coherent quantum coupling between separate wells requires that the distance between ions in neighbouring traps can be reduced in an addressable way \cite{Kumph:2011}. In a two-dimensional array of spherical Paul traps this can be achieved by reducing the power applied to individually adjustable RF electrodes between the traps.

Figure~\ref{fig:ShowingItWorks}a shows the setup for demonstrating RF control of the ion's position. To move an ion away from the home position in the $y$ direction, the power applied to the $y$ adjustable RF electrode was reduced. Figure~\ref{fig:ShowingItWorks}b shows the ion's displacement from the home position as a function of the RF power applied: overall, the ion was displaced by 37~\textmu m for a 5~dB reduction in RF power. For these measurements the phases of the five RF resonators were not locked. As the cooling laser was perpendicular to the displacement direction, any micromotion due to differences in the RF drives' phases would not interfere with laser cooling. Nonetheless, when away from the home position the cooled-ion lifetime was only a few seconds.  This may have been due to a reduction in the cooling power as the ion was displaced, given that the cooling-beam diameter was only 40~\textmu m.

The position of the ion in the $x$ direction was varied using a second, adjustable, phase-locked RF drive, applied to the $x$ adjustable electrode. Again, the ion could be displaced $\sim$40~\textmu m by reducing the applied RF power by $\sim$5~dB on the relevant electrode. For this measurement the resonator was phase locked in order to minimize micromotion along the direction of the cooling beam. Used in combination, the two adjustable drive circuits - one on each of the $x$ and $y$ adjustable electrodes - allowed the ion to be moved ``around a corner" in the trap array.

\subsection{RF tuning of the motional frequency}

In addition to reducing the inter-ion spacing, the interaction between ions neighbouring traps can be increased by lowering their motional frequency (see equation~\ref{eq:GateTime}). As with the reduction of the inter-ion spacing, this can be achieved by reducing the power applied to the relevant adjustable electrode. 

\begin{figure}[t]
\includegraphics[width=135mm]{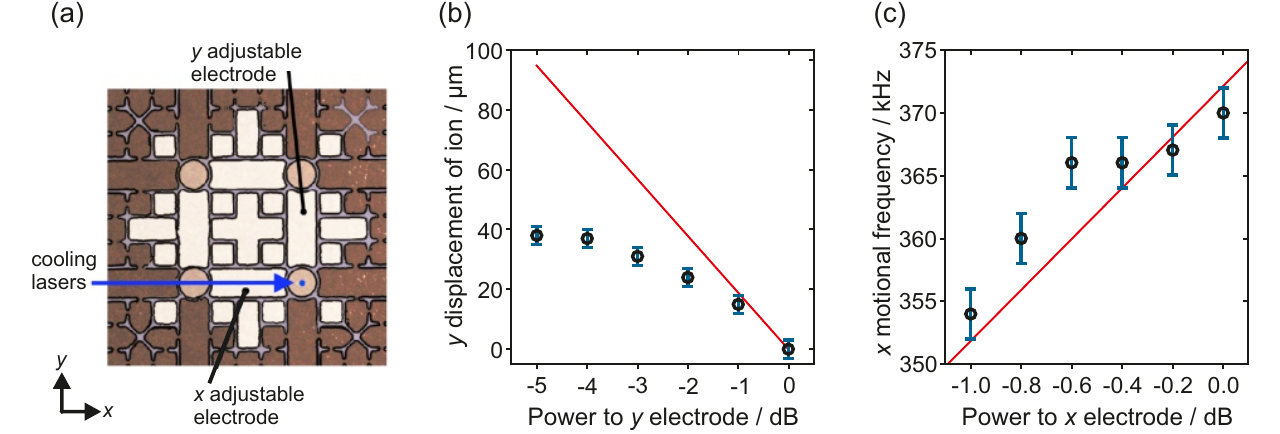}
\caption{Results of varying the power applied to the adjustable electrodes.  (a) Configuration of the laser-cooling and electrode adjustment. The blue dot marks the ion's home position. (b) Points show the displacement of the ion from the home position as a function of the RF power applied to the $y$ electrode. The line shows the simulated position of the RF null. (c) Points show the measured motional frequency of the ion as a function of the RF power applied to the $x$ electrode. The line shows the simulated change in motional frequency.}
\label{fig:ShowingItWorks}
\end{figure}

Figure~\ref{fig:ShowingItWorks}a shows the setup for demonstrating RF control of the ion's frequency. To reduce the motional frequency in the $x$ direction, the power applied to the $x$ adjustable RF electrode was reduced. Figure~\ref{fig:ShowingItWorks}c shows the ion's motional frequency as a function of the RF power applied: overall, the motional frequency was reduced by 15~kHz for a 1~dB reduction in RF power. For these measurements the adjustable RF resonator was phase-locked to the other resonators so that the variation of the RF voltage did not induce excess micromotion. 

The reduction of the applied RF power in the instance shown (as well as the concomitant reduction in the ion's motional frequency) was rather modest. This was because the degradation of the trap behaviour precluded stable trapping for lower RF powers. Operation of the trap at earlier times (with less-severe degradation) demonstrated that stable trapping with applied powers as low as -6~dB is possible (cf. figure~\ref{fig:ShowingItWorks}b). Simulations show that this would correspond to a 25\% reduction in the ions' motional frequency. The limited range of the power adjustment is therefore related to a materials issue of the PCB trap; it is not a limitation of the electronics being demonstrated here, nor is it inherent to the scheme of addressing considered.

\subsection{Heating rate}

The coherent exchange of quantum information between wells in the Cirac-Zoller and the Bermudez schemes motivating this work takes place via the ions' quantized motion. High-fidelity operations require that the ions' motional states are not perturbed by electric-field fluctuations during the time of the gate operation. Such disturbances of the ions' motion are characterized by the ion heating rate, the causes of which are many and varied \cite{Brownnutt:2014}. The heating rates in the current system were measured at two trapping sites: the outer trapping site used in the first trap array, and the inner trapping site used in the second trap array.

At the outer trapping site the heating rate was estimated by an ion-loss method: A single ion was trapped, and the cooling lasers were turned off for increasing lengths of time. For a 5~s waiting time the ion-loss probability was $\sim$50\%. For the stated operating parameters, simulations give a trap depth of $\sim$0.1~eV. By calculating the expected loss-rate from a harmonic trap of this depth for a given (constant) heating rate, assuming a thermal distribution of the ion's motion, a heating rate of $\sim$200~K/s was inferred. In reality the heating rate is likely to increase as the ion becomes hotter, due to the effect of trap anharmonicities far from the trap centre. Additionally, the stated result assumes that the initial ion temperature was much less than the final temperature, which may not have been the case as the ion was not well cooled in the $z$ direction. Consequently the inferred heating rate at this site should be taken as an upper limit of the heating behaviour for cold ions in this trap.

To obtain a more accurate result at the inner trapping site in the second trap array tested, the fluorescence recooling method \cite{Wesenberg:2007} was used to measure the heating rate. The measurements were performed around one year after the trap had first been installed (cf. figure~\ref{fig:trappingvoltage}), by which time the trap behaviour had significantly degraded from its initial condition. Consequently, the uncooled lifetime of ions at this point was only $\sim$100~ms, compared to $\sim$1~s in the first trap array.

\begin{figure*}[t]
\includegraphics[width=130mm]{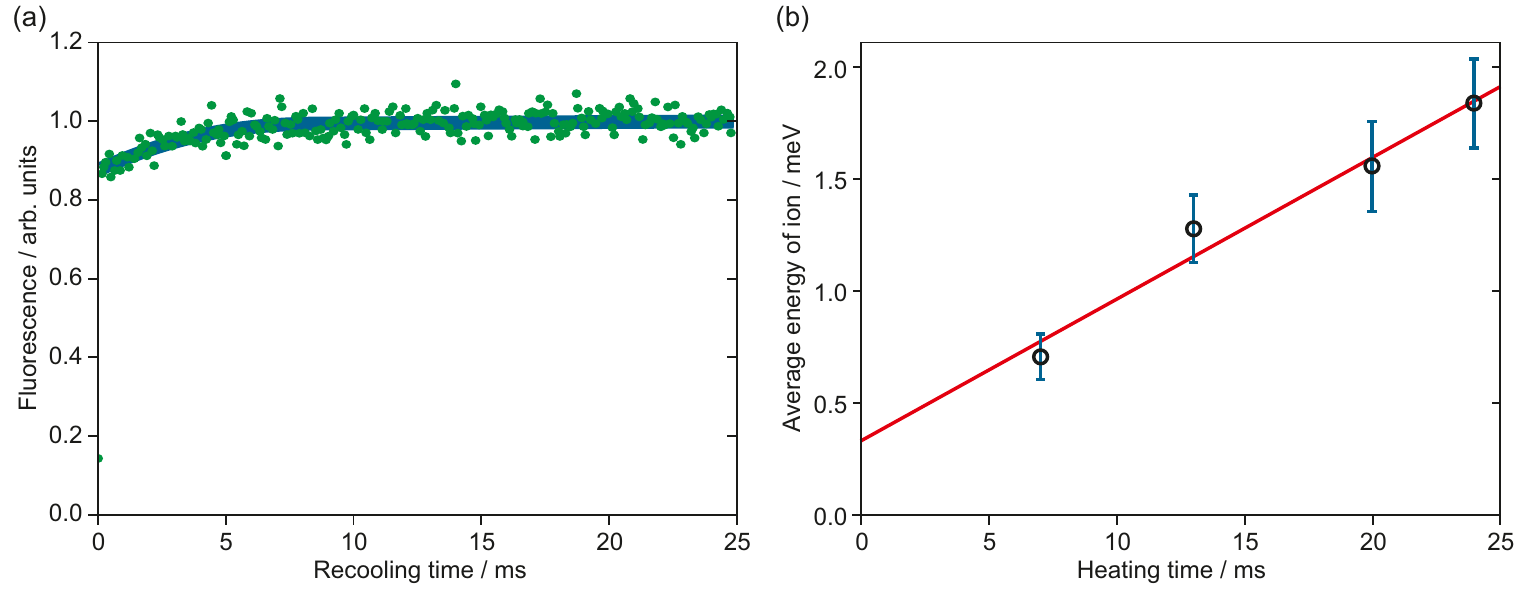}
\caption{Heating-rate measurement using the recooling method. (a) Fluorescence rate as a function of recooling time (point data, showing an average of 1000 experimental runs) and the recooling-model fit (solid line), after allowing the ion to heat for 24~ms. (b) Energy of the ion, given by the recooling model, as a function of the heating time. A linear fit (solid line) gives a heating rate of 60$\pm$20~meV/s.}
\label{fig:HeatingRate}
\end{figure*}

For the heating-rate measurements at the inner trapping site, single \caion ions were loaded by setting all DC voltages (including DC biases on RF electrodes) to zero, except for the relevant trapping-site electrode which, in this instance, was required to be at $-5.3$~V. This large negative voltage increased the ion's radial motional frequency to 630~kHz, but reduced its axial frequency to $\sim$50~kHz. This was a significant perturbation given that, in the absence of DC biases, the radial motional frequency of a spherical Paul trap is half the axial frequency. The ion was Doppler cooled by a laser beam oriented as shown in figure~\ref{fig:ShowingItWorks}a, which had a slight angle ($\sim$6$^\circ$) to the plane of the trap surface in order to cool the axial motion. The cooling laser was then turned off for a certain time, during which the uncooled ion was allowed to heat up. Because the axial frequency was so low compared to the radial frequency, it is assumed that predominantly the axial motion was heated \cite{Brownnutt:2014}.
Following a given ``heating time" the ion was re-illuminated by Doppler-cooling light. The motional state of the ion at the end of the heating time was inferred by analyzing time-resolved measurements of the fluorescence dynamics as the ion was recooled \cite{Wesenberg:2007}. Figure~\ref{fig:HeatingRate}a shows the fluorescence rate during recooling of an ion after it was allowed to heat for 24~ms. For these measurements the 397~nm cooling laser was red-detuned by 10-20~MHz from the centre of the transition, which has a natural linewidth of 21~MHz. The saturation parameter calculated from the measured laser intensity and the linewidth of the effective two-level system was $s=10\pm2$. Similar measurements were carried out for different heating times, shown in figure~\ref{fig:HeatingRate}b. Fitting these points with a line gives a heating rate of (60$\pm$20)~meV/s. This corresponds to (700$\pm$200)~K/s, and is consistent with the short uncooled ion lifetime observed of $\sim$100~ms. The level of electric-field noise required to cause such heating is around $10^{-7}$~$\mathrm{V^2/m^2 Hz}$ \cite{Brownnutt:2014}.

The level of electric-field noise observed in this trap is high, but not unprecedented \cite{Labaziewicz:2008_01}.
The high heating rate does not appear to be caused by the operation of multiple RF electrodes, since the heating rate is also high with only a single RF drive. We conjecture that the high heating rate may be related to the materials used to fabricate the trap, namely gold-plated copper on a PCB substrate. This work provides the first published heating rate in such a trap. This result, along with experience of other copper-on-PCB traps \cite{Splatt:2009} suggests to us that PCB technologies may not be conducive to the production of traps with low heating rates. In light of these considerations, the heating rate---while problematic in this instance---does not appear to be a limitation inherent to the 2D array architecture or to techniques using variable RF.

\section{Outlook and Summary}
\label{sec:DiscussionAndOutlook}

This paper has considered the technological building blocks required to create a 2D array of ion traps in which pairwise interactions between nearest-neighbouring ion traps can be addressably turned on and off. At a single trapping site within a 2D array of ion traps we have demonstrated that an ion's position (in 2D) and its motional frequency can be controllably varied in real time by varying the RF voltage amplitude applied to specific segmented electrodes, while keeping all RF frequencies and phases locked. 

Physically interesting systems can be simulated on relatively small arrays, totalling only a few tens of qubits \cite{Shi:2013, Nielsen:2013}. Section~\ref{subsec:ziegelstadl} discusses the possibilities for miniaturizing the system, so that the distances and frequencies involved are commensurate with demonstrating coherent quantum interactions. Section~\ref{subsec:OutlookScalingUp} considers the prospects for scaling the trap structures, optics and driving electronics to an array of order $10 \times 10$ ions.

\subsection{Miniaturization}
\label{subsec:ziegelstadl}

In order for ions in separate harmonic traps to coherently exchange quantum motion, it is desirable to achieve well separations of $\lesssim$100~\textmu m \cite{Wilson:2014, Kumph:2011}. The PCB prototyping methods used for Folsom are relatively simple at the present scale. However, the minimum possible feature sizes are limited to $\gtrsim$50~\textmu m, which limits the inter-well spacing to $\gtrsim$200~\textmu m. PCB technology therefore cannot be used to produce a viable 2D array of ion traps for quantum information processing. The limitations of PCB technology do not, however, rule out the possibility of realizing a miniaturized architecture using different materials.

\begin{figure}[t]
\includegraphics[width=130mm]{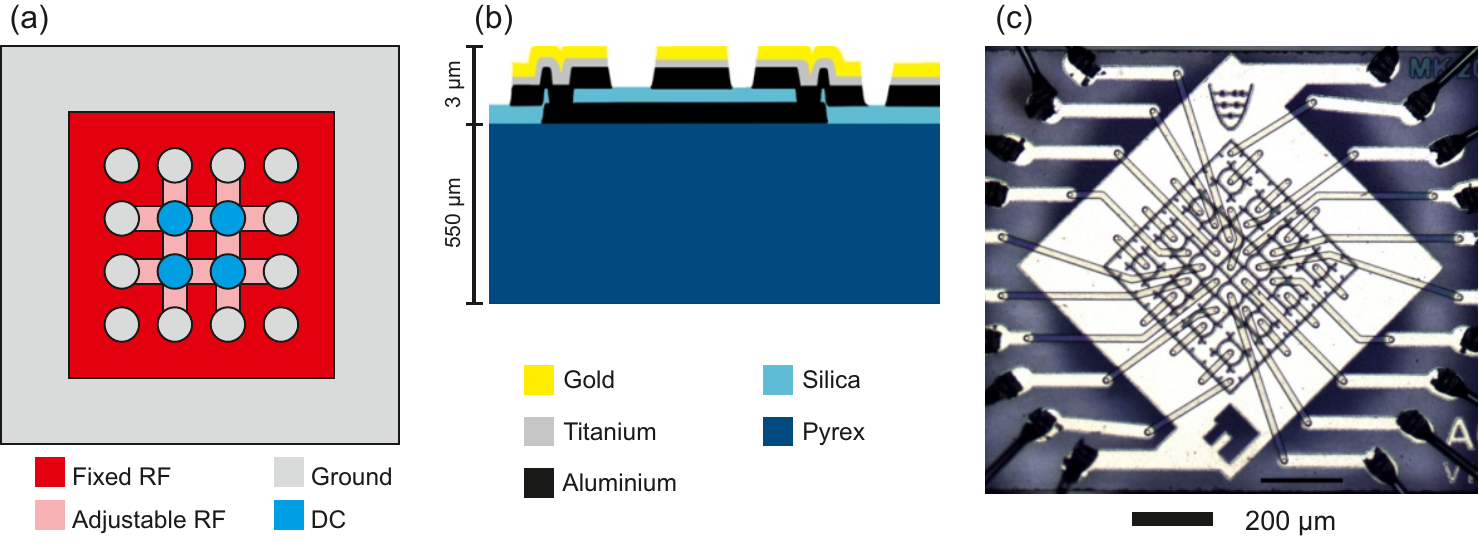}
\caption{Miniaturized trap arrays. (a) Schematic of a simple array design. This requires multi-layer fabrication processes (b) to connect the ``island" trap electrodes to the bond pads at the edges of the array. (c) Prototype trap currently under test.}
\label{fig:Ziegelstadl}
\end{figure}

A simple electrode geometry (shown in figure~\ref{fig:Ziegelstadl}a) can be considered in which the distance between the trapped ions in their home position is 100~\textmu m, and for which the ions are 50~\textmu m above the surface. Varying the voltage on the adjustable RF electrodes would allow the distance between a pair of ions to be reduced to 60~\textmu m during an interaction, while maintaining distinct potential wells. A trap drive with a voltage amplitude of 150~V and a frequency of 100~MHz would give a motional frequency of 10~MHz and a trapping-potential depth of 0.1~eV. This would lead to a gate time, $T_{\rm gate}$, of a few ms. This gate time compares favourably to the heating rate which may reasonably be achievable in a trap of this size - with care and at cryogenic temperatures -  of a few phonons per second \cite{Brownnutt:2014}.

The size scales and parameter ranges outlined above are within reach of current fabrication methods \cite{Guise:2015}. Fabrication efforts to realize a 2D array of ion traps which is at a scale suitable for quantum information processing are currently underway. The traps are fabricated from aluminium on Pyrex with a gold/titanium surface coating (shown in figure~\ref{fig:Ziegelstadl}b). Prototype trap arrays (shown in figure~\ref{fig:Ziegelstadl}c) are currently under test.

\subsection{Larger numbers of qubits}
\label{subsec:OutlookScalingUp}

The complexity of the electrode layout and driving electronics increases linearly with the number of trapping sites: the basic designs shown in figures \ref{fig:ShowingItWorks}a and \ref{fig:phaselock} can simply be replicated many times over. A square array of $N \times N$ ions requires $\sim 2 N^2$ independent RF voltage sources. For $N=10$ this is possible using the same surface-mount components as were used for the present work. Alternatively, integrated electronic components may be used \cite{Guise:2015} allowing significant miniaturization of a number of components. At room temperature and at trap-drive frequencies of 100~MHz it is not trivial to reduce the physical size of the inductors in the resonators. The use of higher drive frequencies allows smaller inductors to be used. Also operation at cryogenic temperatures allows the possibility of using very much smaller kinetic inductors \cite{Meservey:1969}.

As larger arrays are used, a more involved optical setup will be required. To cool the ions it would be useful to have a laser beam broad enough (in the $x$ direction) to cool all ions simultaneously, while being sufficiently tightly focused (in the $z$ direction) to not significantly illuminate (and thereby cause charging of) the trap electrodes. A highly elliptical light sheet which is $\sim 1$\,mm in one direction, while being focused to a waist of $w_0 \approx 10$\,\textmu m in the other can be achieved for all relevant wavelengths for Ca$^+$ ions using standard optics.

\subsection{Summary}

Two-dimensional arrays of ions in which interactions can be individually addressed constitute an attractive architecture for trapped-ion quantum computing. We have demonstrated the operation of 2D arrays of surface-electrode point ion traps in which the power supplied to specific RF-electrode segments can be varied, while keeping the phase, frequency and all other RF-voltage amplitudes constant. By this means we have demonstrated that an ion's position can be changed and its motional frequency tuned in two dimensions. This lays the foundations for addressable, tunable, nearest-neighbour interactions between trapped ions in a 2D array.

\section*{Acknowledgements}

The miniaturized trap array shown in Fig.~\ref{fig:Ziegelstadl}, and dubbed ``\textit{Ziegelstadl}", was fabricated by S. Partel in the research group of Prof. Edlinger, Fachhochschule Vorarlberg, Dornbirn.
This work was supported by
the European Research Council (ERC) through the Advanced Research Project CRYTERION
and the Proof of Concept Project CARAT;
the Austrian Science Fund (FWF) project Q-SAIL;
and the Institute for Quantum Information GmbH.

\newpage


\end{document}